\documentclass
[superscriptaddress,secnumarabic,nobibnotes,aps,prd,showkeys,noshowpacs,twocolumn,10pt]{revtex4-2}%
\usepackage{graphics}
\usepackage{graphicx}
\usepackage{subcaption}
\usepackage{epsf}
\usepackage{bm}
\usepackage{amsmath,amssymb,amsfonts,mathrsfs,amsthm}
\usepackage{latexsym}
\usepackage{enumerate}
\usepackage{comment}
\usepackage[dvipsnames,svgnames,x11names,table]{xcolor}
\usepackage[colorlinks = true,
            linkcolor = Cerulean,
            urlcolor  = magenta,
            citecolor = magenta,
            anchorcolor = NavyBlue]{hyperref}
\usepackage{epstopdf}%
\usepackage{bbm}
\def\be{\begin{equation}}
\def\ee{\end{equation}}

\newcommand{\ima}{\mathrm{i}}
\makeatother

\begin{document}

\title{Time operator from parametrization invariance and implications for cosmology}
\author{N. Dimakis}
\email{nikolaos.dimakis@ufrontera.cl}
\affiliation{Departamento de Ciencias F\'{\i}sicas, Universidad de la Frontera, Casilla 54-D, 4811186 Temuco, Chile}

\begin{abstract}\noindent
  Motivated by the parametrization invariance of cosmological Lagrangians and their equivalence to systems describing the motion of particles in curved backgrounds, we identify the phase space analogue of the notion of proper time. We define the corresponding quantum operator, which results in being canonically conjugate to that of the vanishing Hamiltonian. In the context of particle dynamics, this leads to an uncertainty relation of the form $\Delta E_0 \,\Delta T \geq \hbar$, where $E_0$ is the rest energy of the particle. By studying the non-relativistic limit, we show that the action of the operator reduces to multiplication by the classical time coordinate. Finally, we derive the generic expression for the introduced time variable in the cosmological setting.
\end{abstract}

\maketitle

\section{Introduction}

In the quest for seeking a quantum description of gravitational theories many different paths are actively explored \cite{QG1,QG2,QG3,QG4,QG5,QG6,QG7,QG8,QG9,QG10}. Cosmological configurations \cite{QC1,QC2,QC3,QC4,QC5,QC6,QC7,QC8,QC9,QC10,QC11,QC12,QC13,QC14,QC15,QC16,QC17,QC18,QC19,QC20,QC21,QC22,QC23,QC24,QC25,QC26,QC27,QC28} and black hole spacetimes \cite{BH1,BH2,BH3,BH4,BH5,BH6,BH7,BH8,BH9,BH10,BH11,BH12,BH13,BH14} offer the simplest possible gravitational settings which are investigated with respect to their quantum properties. 

One of the most fundamental challenges in the attempt to reconcile gravity with the quantum theory is the well-known ``problem of time'' \cite{PT1,PT2,PT3,PT4,PT5,PT6,PT7,PT8,PT9,PT10,PT11,PT12}. Quantum Mechanics (QM) was initially developed for regular systems, where time is an absolute parameter. The parametrization invariance of cosmological Lagrangians, which results in the Hamiltonian being zero, generates severe interpretational issues on how time evolution should be perceived. A common approach is to adopt an internal degree of freedom as an effective time variable, with respect to which the rest of the dynamical quantities are expressed \cite{IT1,IT2,IT3,IT4,IT5}. Other approaches of introducing a time operator in parametrization invariance systems can be found in \cite{Deriglazov,Nandi1,Nandi2}.

In this work, we exploit the analogy between cosmological Lagrangians and those describing the motion of particles in curved spaces, i.e. geodesic Lagrangians \cite{geo1,geo2,geo3,geo4,geo5,geo6,geo7}. From the geometric properties of the underlying manifold, a phase-space function emerges, which naturally assumes the role of proper time. By following the canonical quantization procedure, we construct the corresponding quantum operator and prove that it satisfies the expected canonical commutation relation with the Hamiltonian. Applying this framework to the Klein-Gordon equation, we demonstrate that in the non-relativistic limit, the operator action reduces to a simple multiplication with the time coordinate. For spatially homogeneous cosmologies, we derive the expression of the newly introduced time with respect to the scale factors. Finally, we briefly demonstrate the validity of the derived relations through a simple example in Friedmann–Lema\^itre–Robertson–Walker (FLRW) cosmology.
%and we finalize our analysis with some concluding remarks. 

\section{Classical description}

Cosmological Lagrangians, after the process of a mini-superspace reduction, assume the form
\begin{equation} \label{cosmoLag}
L = \frac{\mathcal{V}_0}{2 N} \bar{G}_{\mu\nu}(q) \dot{q}^\mu \dot{q}^\nu - N \mathcal{V}_0 V(q),
\end{equation}
where $N$ is the lapse function, $\bar{G}_{\mu\nu}$ is the mini-superspace metric and the $q^\mu$ denote the gravitational and matter degrees of freedom. The $\mathcal{V}_0$ is an overall constant representing the three-dimensional volume; a remnant of integrating out the spatial degrees of freedom during the mini-superspace approximation \cite{vol1,vol2}. 

A reparametrization of the lapse function, $N\rightarrow n = N V(q)/\mathcal{V}_0$, leads to
\begin{equation} \label{Lag}
L = \frac{1}{2 n} G_{\mu\nu} (q) \dot{q}^\mu \dot{q}^\nu - n \frac{ m^2 c^2}{2}, 
\end{equation}
where we have set $\mathcal{V}_0^2=(mc)^2/2$. Lagrangian \eqref{Lag} describes the motion of a free particle of mass $m$ in a curved space with metric $G_{\mu\nu} = V \, \bar{G}_{\mu\nu}$. The $n$ in this context plays the auxiliary role of the einbein field \cite{einb}. Lagrangians \eqref{cosmoLag} and \eqref{Lag} are parametrization invariant, meaning that the parameter of the particle trajectory, say $\tau$, can be changed in an arbitrary manner. The dots over the $q$'s in \eqref{cosmoLag} and \eqref{Lag} represent derivation with respect to this parameter. 

The geometric symmetries of $G_{\mu\nu}$ are related to conservation laws \cite{Katz,Andr,tchris1,Tsamp}. Specifically, the conformal Killing vectors (CKVs) of $G_{\mu\nu}$, which satisfy 
\begin{equation} \label{homeq}
  \nabla_\mu \xi_\nu + \nabla_\nu \xi_\mu =2 \omega(q) G_{\mu\nu},
\end{equation}
generate conserved charges \cite{tchris1,geo5}. The general form of the aforementioned conserved charges is nonlocal, since it involves an integral over the degrees of freedom \cite{dim1}
\begin{equation} \label{inthom}
  I  = Q(q,p) + m^2 c^2 \int \!\! n(\tau)   \omega(q(\tau))  d\tau .
\end{equation}
In certain cases, these are shown to be associated with the existence of hidden symmetries \cite{Horv,dim1,dim2}. The $Q(q,p)$ in the above expression is a linear in the momenta quantity, 
\begin{equation}
  Q(q,p)=\xi^\mu p_\mu ,
\end{equation}
with $\xi^\mu$ being the components of the vector satisfying the conformal Killing equation \eqref{homeq}. Note that for zero-mass particles, the conserved charge reduces to the well known expression, $I=\xi^\mu p_\mu$, for any CKV. The same is true if $\xi$ is a Killing vector of $G_{\mu\nu}$, i.e. if $\omega=0$, then again Eq. \eqref{inthom} reduces to the well-known $I=\xi^\mu p_\mu$.

The conservation of the integral of motion given in Eq. \eqref{inthom} can be easily verified by considering the Hamiltonian associated to Lagrangian \eqref{Lag}, which is
\begin{equation} \label{clasHam}
  H = \frac{n}{2} \mathcal{H} + u_n p_n .
\end{equation}
The $u_n$ is an arbitrary multiplier, which accounts for the missing velocity $\dot{n}$ in the Lagrangian. The $p_n := \frac{\partial L}{\partial \dot{n}} \approx 0$ is the primary constraint and
\begin{equation}\label{Hamcon}
  \mathcal{H} = G^{\mu\nu} p_\mu p_\nu +  m^2 c^2 \approx 0
\end{equation}
is the Hamiltonian (or quadratic) constraint. The fact that the Hamiltonian is zero is a direct consequence of the underlying parametrization invariance of the system. Returning to the derivation of the conservation law, notice that by taking the total derivative of $I$ we obtain
\begin{equation}
  \frac{d I}{d\tau} = \frac{\partial I}{\partial \tau} + \{I,H\} = n\, \omega\, \mathcal{H} \approx 0.
\end{equation}
Thus, $I$ is conserved due to the fact that the Hamiltonian is zero. We use the symbol ``$\approx$'' in the Dirac sense of a weak equality \cite{Dirac}, meaning that the equality holds after the enforcement of the constraints - a process that takes place only when the constraints appear out of Poisson brackets \cite{Sund}. 

Notice that in the gauge $n=$const., which converts $\tau$ into the proper time of the particle, if $\xi$ represents a homothetic vector, i.e. $\omega=$const., then the integral term in $I$ becomes a function linear in $\tau$. For example, setting $n=1$, $\omega=1$ in Eq. \eqref{inthom} yields
\begin{equation} \label{propercl}
  I = \xi^\mu p_\mu + m^2 c^2   \tau ,
\end{equation}
where $\tau$, as already mentioned, represents now the proper time. Given that $I$ is a constant of the motion, Eq. \eqref{propercl} suggests that we can identify the time flow with a quantity proportional to $\xi^\mu p_\mu$. In particular, note that 
\begin{equation} \label{cc1}
  \{-\xi^\mu p_\mu/mc^2,\mathcal{H}/m\} \approx 1 .
\end{equation}
This linear in the momenta quantity, which we constructed with the help of the homothetic symmetry, is canonically conjugate to the Hamiltonian constraint. The above equality holds by virtue of the fact that the Hamiltonian constraint is zero, hence the use of ``$\approx$''. Our purpose is to use this classic relation as a guide to help us identify a reasonable time operator in the quantum description of the system.

\section{Quantization}

In the process of canonical quantization, and in order to address the factor ordering problem of the kinetic term in \eqref{Hamcon}, a common choice, especially in the cosmological setting, is the conformal Laplacian (or Yamabe operator) \cite{tchris2,Ryan,Anderson},
\begin{equation} \label{Yam}
\widehat{L}_g = \frac{1}{\mu} \partial_\alpha \left(\mu \, G^{\alpha\beta} \partial_\beta \right) - \frac{d-2}{4 (d-1)} R,
\end{equation}
where $d = \dim{G_{\alpha\beta}}$ is the dimensionality of the (mini-superspace) metric, $\mu=\det{|G_{\mu\nu}|}$ is the natural measure obtained from the determinant, and $R$ is the scalar curvature of $G_{\mu\nu}$. With this choice, the quadratic constraint operator reads
\begin{equation} \label{conop}
\widehat{\mathcal{H}} = - \hbar^2 \widehat{L}_g + m^2 c^2.
\end{equation}
This is a Hermitian operator under the measure $\mu$ and for appropriate boundary conditions for the wave function. Returning to the linear in the momenta part of $I$ in \eqref{inthom} denoted with $Q=\xi^\mu p_\mu$, we may write the corresponding operator as
\begin{equation} \label{linop}
  \begin{split}
    \widehat{Q} &  = - \frac{\ima \hbar}{2 \mu} \left( \mu\, \xi^\alpha \partial_\alpha + \partial_\alpha \left(\mu \, \xi^\alpha \right)\right) \\
    & = -\ima \hbar \xi^\alpha \nabla_\alpha - \frac{\ima \hbar}{2} \nabla_\alpha \xi^\alpha .
  \end{split}
\end{equation}
The above is the more general linear, first order differential operator with the property of being Hermitian under the measure $\mu$  \cite{tchrisop,tchrisBH}. 

We shall now make use of the symmetries of the conformal Laplacian, which are well studied in \cite{Eastwood,CL1}. According to theorem 1 of \cite{Eastwood}, whenever $\xi$ is a conformal Killing vector field of $G_{\mu\nu}$, the operators defined as
\begin{subequations} \label{defopsd}
\begin{align}  \label{defD}
\widehat{D}_\xi & = \xi^\alpha \nabla_\alpha + \frac{d-2}{2d}  \left( \nabla_\alpha \xi^\alpha \right) \\ \label{defd}
\widehat{\delta}_\xi & = \xi^\alpha \nabla_\alpha + \frac{d+2}{2d} \left( \nabla_\alpha  \xi^\alpha \right) ,
\end{align}
\end{subequations}
satisfy
\begin{equation} \label{symcon}
  \widehat{L}_g \widehat{D}_\xi = \widehat{\delta}_\xi \widehat{L}_g .
\end{equation}

Considering the action of \eqref{linop} and \eqref{defopsd} on a function $\Psi$ it is clear that we can write
\begin{align} \nonumber
-\ima \hbar \, \widehat{D}_\xi \Psi & = \widehat{Q} \Psi + \frac{\ima \hbar}{d} \left(\nabla_\alpha \xi^\alpha \right) \Psi \\
-\ima \hbar \, \widehat{\delta}_\xi \Psi & = \widehat{Q} \Psi - \frac{\ima \hbar}{d} \left( \nabla_\alpha \xi^\alpha \right) \Psi
\end{align}
As a result, Eq. \eqref{symcon} becomes
\begin{equation} \label{interim1}
-\hbar^2 \widehat{L}_g \left( \widehat{Q}\Psi + \frac{\ima \hbar}{d} \left( \nabla_\alpha \xi^\alpha \right) \Psi \right) = -\ima \hbar \, \widehat{\delta}_\xi (-\hbar^2\widehat{L}_g \Psi).
\end{equation}

We now proceed by making the following assumptions:
\begin{enumerate}
  \item $\Psi$ is a wave function satisfying Dirac's quantization prescription for systems with constraints, that is
  \begin{equation}
    \widehat{p}_n \Psi = -\ima \hbar \frac{\partial\Psi}{\partial n} =0 \quad \text{and} \quad \widehat{\mathcal{H}} \Psi =0
  \end{equation}
  The first condition implies that $\Psi$ is independent of $n$. The second, with the help of Eq.  \eqref{conop}, leads to $\hbar^2\widehat{L}_g \Psi = m^2 c^2 \Psi$. In the case of cosmology this is the Wheeler-DeWitt equation (WDW).
      
  \item $\xi$ is a homothetic vector to the metric $G_{\mu\nu}$. From the definition of conformal Killing vectors \eqref{homeq} we know that $\nabla_\alpha \xi^\alpha = \omega\, d$. For a homothetic vector this is a constant, and without loss of generality we can set $\omega=1$, so that $\nabla_\alpha \xi^\alpha = d$. 
\end{enumerate}

With these assertions, Eq. \eqref{interim1} leads us to
\begin{equation} \label{interim2}
  \begin{split}
     -\hbar^2\widehat{L}_g \left( \widehat{Q}\Psi + \ima \hbar \Psi \right) & = -m^2 c^2 \left(\widehat{Q}\Psi - \ima \hbar  \Psi  \right)\Rightarrow \\ 
     \left(-\hbar^2\widehat{L}_g + m^2 c^2\right) \widehat{Q} \Psi & = 2\ima \hbar\, m^2 c^2 \Psi \Rightarrow \\
     \widehat{\mathcal{H}} \widehat{Q} \Psi & = 2\ima \hbar\, m^2 c^2 \Psi .
  \end{split}
\end{equation}
Since $\Psi$ satisfies the WDW equation, $\widehat{\mathcal{H}} \Psi =0$, we can trivially write the above relation as
\begin{equation} \label{commu1}
  [\widehat{\mathcal{H}}, \widehat{Q} ]\Psi = 2 \ima  \hbar \, m^2 c^2  \Psi .
\end{equation}
This result is in accordance with the classical description, where the time derivative of a phase space quantity $Q(q,p)$ is given on mass shell by $\dot{Q} = \{ Q, H\}$. To see this consider the quantum analogue of Hamiltonian \eqref{clasHam} given by
\begin{equation} \label{Hamquant}
\widehat{H} = \frac{n}{2} \widehat{\mathcal{H}} + u_n (t) \widehat{p}_n, 
\end{equation}
where $u_n$ remains an arbitrary multiplier. A direct application of the canonical quantization rule, $\{\;,\;\} \rightarrow -\frac{\ima}{\hbar}  [\; ,\; ]$, allows us to write
\begin{equation}
\dot{\widehat{Q}} = -\frac{\ima}{\hbar}  [ \widehat{Q},\widehat{H}] .
\end{equation}
The latter, together with Eq. \eqref{commu1}, results in the action of the operator $\dot{\widehat{Q}}$ on the solutions of the WDW equation being
\begin{equation}  \label{qhdot}
\dot{\widehat{Q}} \Psi = - n\, m^2 c^2 \Psi .
\end{equation}
This is in perfect agreement with the classical expression for the integral of motion of Eq. \eqref{inthom}, which in the case of a homothety ($\omega=1$) is simply 
\begin{equation}
  I=Q+m^2 c^2 \int\!\! n \, d\tau
\end{equation}
and since $I$ is a constant of the motion due to the vanishing Hamiltonian, i.e. $\dot{I}\approx 0$, we have
\begin{equation}
  \dot{I}= \dot{Q} + n \, m^2 c^2 \approx 0.
\end{equation}

In the same spirit, and remembering  that the proper time, in the gauge $n=1$, is retrieved from \eqref{propercl} as proportional to $\xi^\alpha p_\alpha$, we define what we call a \emph{quantum time operator}
\begin{equation} \label{timeop}
  \widehat{T} := -\frac{1}{m c^2} \widehat{Q} .
\end{equation} 
It immediately follows from \eqref{commu1} and \eqref{Hamquant} that
\begin{equation}
  [\widehat{T},\widehat{H}/m]\Psi = \ima \hbar n \Psi, 
\end{equation}
which, in the gauge $n=1$ of the proper time, becomes 
\begin{equation} \label{canTH}
  [\widehat{T},\widehat{H}/m] \Psi = \ima \hbar \Psi .
\end{equation}
That is, $\widehat{T}$, is the canonical conjugate of the Hamiltonian, or more precisely of $\widehat{H}/m$, which we can see from Eqs. \eqref{clasHam} and \eqref{Hamcon} that it has units of energy. The classical equivalent of $\widehat{T}$ is of course $T=-Q/(m c^2)$, which we have already seen from Eq. \eqref{cc1} that it satisfies 
\begin{equation}
  \dot{T}=\{T,H/m\}\approx 1 .
\end{equation}
This last relation implies that $T$ is by definition monotonic in this gauge, thus a good variable to be used as time. An important point is to recall that these classical equalities hold by virtue of the constraint $\mathcal{H}\approx 0$. In a sense the same applies at the quantum level as well since, for the above relations to hold, $\Psi$ has to be a solution of the WDW equation. 

\section{Time-energy uncertainty}

In QM it is well established how the uncertainty relation $\Delta q \, \Delta p   \geq \hbar/2$ emanates from the quantum canonical commutator $[\widehat{q}, \widehat{p} \,] = \ima \hbar$. Specifically, for any two linear, Hermitian operators $\widehat{A}$ and $\widehat{B}$, we have that their standard deviations, $\Delta A = (\langle\widehat{A}^2\rangle-\langle\widehat{A}\rangle^2)^{1/2}$ and similarly for $\Delta B$, satisfy \cite{uncert}
\begin{equation}\label{uncgen}
  \Delta A \, \Delta B \geq \frac{1}{2} |\langle[\widehat{A},\widehat{B}]\rangle|.
\end{equation}
In the typical theory of QM, a time-energy uncertainty relation, although usually implied in the literature \cite{unc1}, does not stem from a canonical commutation relation like it happens for the position-momentum uncertainty. This is due to the lack of a time operator and because of the role that time plays in the measuring process \cite{Aharonov}. Pauli even introduced a theorem arguing the impossibility of the task \cite{Pauli}, later developments however, have challenged the mathematical rigor of this theorem and the conditions for its derivation \cite{cPau0,cPau1,cPau2}. According to the initial version of Pauli's argument, there could not exist a Hermitian time operator consistent with a bounded Hamiltonian operator. However, the formulation of Pauli's objection did not take into account the domains of the aforementioned operators. It has later been shown that it is actually possible to define time operators for semi-bounded or even bounded Hamiltonians \cite{cPau3,cPau4}. 

Our definition for the time operator, given in \eqref{timeop}, together with Eq. \eqref{canTH}, is tempting enough to introduce such a relation in a natural manner. In order to show this explicitly, let us define an operator $\widehat{E}_0$ so that $\widehat{\mathcal{H}}/m=-\widehat{E}_0+ mc^2$. In other words, we set $\widehat{E}_0=\hbar^2 \widehat{L}_g/m$. Now, by virtue of the constraint condition, that is the WDW equation $\widehat{\mathcal{H}}\Psi=0$, we have $\widehat{E}_0\Psi =m c^2 \Psi$ on the physical states. So, $\widehat{E}_0$ is an operator associated to the rest energy of the particle. Then, with a simple use of \eqref{commu1} and our definition of $\widehat{T}$ from \eqref{timeop} we obtain the desired time-energy canonical commutation relation
\begin{equation} \label{unc}
  [\widehat{E}_0,\widehat{T}]\Psi = 2 \ima  \hbar \, \Psi,
\end{equation}
thus implying $\Delta E_0 \, \Delta T \geq \hbar$. However, note the interpretation of such a relation here is delicate. Remember that we derived the quantum expressions, like Eq. \eqref{canTH}, under the condition that $\Psi$ satisfies the constraint $\widehat{\mathcal{H}}\Psi=0$. If $m\, c^2$ here is to be considered as a fixed number in the problem, as is usually the case due to the nature of this equation being a constraint, then it does not really make sense to talk of an uncertainty in $\widehat{E}_0$ since it is uniquely fixed. However, if, on the other hand, we interpret $\widehat{\mathcal{H}}\Psi=0\Rightarrow \widehat{E}_0\Psi=m\, c^2\Psi$ as an eigenvalue equation, where $m\, c^2$ may actually run through different values, then, in that case, it is reasonable to consider \eqref{unc} as the starting point for a time-energy uncertainty relation. 

Another implication of the non-commutativity between $\widehat{T}$ and $\widehat{\mathcal{H}}$, which is fairly obvious, but nevertheless deserves mentioning, is that an eigenvalue equation $\widehat{T}\Psi= \kappa \Psi$, where $\kappa=$const. cannot be satisfied by physical states since $\widehat{T}$ and $\widehat{\mathcal{H}}$ cannot be ``measured'' simultaneously. In a sense, the fact that the Hamiltonian of parametrization invariant systems is zero guarantees that time cannot be ``frozen'' in a given system. A solution to the WDW equation cannot have $\widehat{T}$ as an eigenoperator. Note that, the action of $\widehat{T}$ on $\Psi$, takes us off the physical Hilbert space, $\widehat{T}\Psi=:Y(q)$, with $\widehat{\mathcal{H}}Y\neq0$. However, successive use of the Hamiltonian constraint returns us back inside, since from Eq. \eqref{interim2} we obtain $\widehat{\mathcal{H}}Y=\widehat{\mathcal{H}}\widehat{T}\Psi=-2\ima \hbar \Psi$.

\section{The relativistic particle example}

In order to see how the operator $\widehat{T}$ connects with what we know from the usual QM, let us consider the motion of a free particle in Minkowski space 
\begin{equation} \label{Minkowski}
  ds^2 =- c^2 dt^2 + dx^2 + dy^2 + dz^2.
\end{equation}
The Hamiltonian constraint \eqref{Hamcon} in this case is simply
\begin{equation}
  -p_0^2 + p_x^2 + p_y^2 + p_z^2 + m^2 c^2 =0 ,
\end{equation}
which upon substitution of $p_0=E/c$ results in the well known relation $E^2 = p^2 c^2 + m^2 c^4$. 

Quantization according to Dirac's prescription leads to the Klein-Gordon equation
\begin{equation}\label{KG}
   \left(\frac{1}{c^2}\frac{\partial^2}{\partial t^2} - \nabla^2 + \frac{m^2 c^2}{\hbar^2}\right) \Psi =0.
\end{equation}
As is known, the non-relativistic limit leads to the Schr\"odinger equation. To see this consider the ansatz $\Psi = e^{-\ima m c^2 t/\hbar}\psi(t,x,y,z)$, which assigns part of the time dependence in the wave function to the rest energy $E_0=m c^2$. In the non-relativistic limit this energy dominates the total energy of the particle. Then, by substitution in \eqref{KG} and by depreciating terms divided by $c^2$, the result is \cite{Greiner}
\begin{equation}\label{Sch}
  \ima \hbar \frac{\partial \psi}{\partial t} = - \frac{\hbar^2}{2m} \nabla^2 \psi .
\end{equation}

Let us perform the same non-relativistic reduction over the result of the operator $\widehat{T}$ on $\Psi$. First notice that the homothetic vector of the metric of the space where the motion takes place is given by
\begin{equation}
  \xi = t\frac{\partial}{\partial t} + x^i \frac{\partial}{\partial x^i}.
\end{equation}
It is easy to verify that for the Minkowski metric \eqref{Minkowski} the above vector satisfies Eq. \eqref{homeq} with $\omega=1$. The classical $T$ quantity is then given by $T=-\left( t\, p_t +x^i p_i\right)/(m c^2)$, while its quantum analogue, according to \eqref{timeop} and \eqref{linop}, reads
\begin{equation} \label{timeopex}
  \widehat{T} = \ima \frac{\hbar}{m\, c^2} \left( t\frac{\partial}{\partial t} + x^i \frac{\partial}{\partial x^i} + \frac{1}{2}\right).
\end{equation}
Using the same ansatz for $\Psi$ as before we obtain 
\begin{equation}
  \begin{split}
    \widehat{T}\Psi & = e^{-\frac{\ima m c^2 t}{\hbar}}\left[t \psi + \frac{\ima \hbar}{c^2} \left( t \frac{\partial \psi}{\partial t} + x^i \frac{\partial \psi}{\partial x^i} +\frac{\psi}{2} \right) \right]\Rightarrow \\
    \widehat{T}\Psi & \simeq t \Psi ,
  \end{split}
\end{equation}
where in the last approximate equality we eliminated the terms dived by $c^2$ in the same manner we did in the derivation of \eqref{Sch}. It is striking that the action of $\widehat{T}$ in this limit reduces to a simple multiplication with the time variable. In the non-relativistic limit $\widehat{T}$ turns basically into what is perceived as a $c$-number. 

\section{Cosmology and connection to York time}

Up to now we considered the time operator from the standpoint of geodesic motion. It would be interesting to study the analogy from a cosmological perspective. In General Relativity (GR) Lagrangians like \eqref{cosmoLag}, derived in the case of spatially homogenous spacetimes in vacuum, admit a mini-superspace metric of the form \cite{miniform}
\begin{equation}
  \bar{G}^{ijkl} = \frac{1}{4} \sqrt{\gamma}\left( \gamma^{ik}\gamma^{jl} + \gamma^{il}\gamma^{jk} - 2 \gamma^{ij}\gamma^{kl} \right) .
\end{equation}
The Lagrangian itself reads
\begin{equation} \label{minioriginal}
  L = \frac{\mathcal{V}_0}{2\kappa N} \bar{G}^{ijkl} \dot{\gamma}_{ij} \dot{\gamma}_{kl} + N \frac{\mathcal{V}_0}{\kappa} \sqrt{\gamma} \; {}^{(3)}\! R,
\end{equation}
where $\kappa=8\pi G_N/c^4$ is the gravitational constant. We include this constant here, while we did not do so when writing \eqref{cosmoLag}. The reason is that Lagrangian \eqref{cosmoLag} is meant to incorporate all possible degrees of freedom including matter fields. In such a case, $1/\kappa$ will not appear multiplicatively in front of every term in the Lagrangian. 

The $\gamma_{ij}$ in \eqref{minioriginal} denote the entries of the scale factor matrix, $\gamma$ is its determinant and ${}^{(3)}\!R$ the spatial curvature of the three-metric. We thus have the correspondence $\bar{G}_{\mu\nu} \sim G^{ijkl}$, $V \sim \sqrt{\gamma} \; {}^{(3)}\! R$ with the coordinates being $q^\mu = \gamma_{ij}$. The geodesic equivalent problem is described by
\begin{equation} \label{miniogeo}
  L_g = \frac{1}{2\kappa n }\sqrt{\gamma} \; {}^{(3)}\! R\, \bar{G}^{ijkl} \dot{\gamma}_{ij} \dot{\gamma}_{kl} + n \frac{\mathcal{V}_0^2}{\kappa},
\end{equation}
where we introduced $n=N \sqrt{\gamma} \; {}^{(3)}\! R  /\mathcal{V}_0$.

Remarkably, the $\bar{G}^{ijkl}$ and the potential $V \sim \sqrt{\gamma} \; {}^{(3)}\! R$ admit the same vector as a homothety, which means that also the scaled mini-superspace metric 
\begin{equation}
  G^{ijkl}=\sqrt{\gamma} \; {}^{(3)}\! R \, \bar{G}^{ijkl}
\end{equation}
of the equivalent geodesic problem has the same homothety. It is easy to see that this vector is the generator of scalings in the $\gamma_{ij}$, i.e. $\gamma_{ij}\frac{\partial}{\partial \gamma_{ij}}$. To see this consider a scaling transformation $\gamma_{ij}\rightarrow \lambda \gamma_{ij}$. Then the mini-superspace scales as $\bar{G}^{ijkl}\rightarrow \lambda^{\frac{d}{2}-2}\bar{G}^{ijkl}$ and the potential as $V\rightarrow \lambda^{\frac{d}{2}-1} V$, where $d$ is the number of the $\gamma_{ij}$'s, i.e. the dimension of the mini-superspace. As a result, $G^{ijkl}$ scales as $G^{ijkl}\rightarrow \lambda^{d-3}G^{ijkl}$ \cite{tchris1}.

The classical analogue of our quantum time operator is proportional to the phase space quantity $\xi^\mu p_\mu$, which in this case is, up to some multiplicative constant, given by
\begin{equation} \label{Tmini}
  T \sim \gamma_{ij} \pi^{ij},
\end{equation}
where $\pi^{ij}= \frac{\partial L_g}{\partial \dot{\gamma}_{ij}}$ are the momenta. This is up to an overall multiplicative factor reminiscent of the York time \cite{York1,York2}. The latter is defined in terms of the extrinsic curvature of the spatial slices and assumes the expression
\begin{equation}
  T_Y = \frac{2}{3\sqrt{\gamma}} \gamma_{ij} \bar{\pi}^{ij} ,
\end{equation}
where the $\bar{\pi}^{ij}=\frac{\partial L}{\partial \dot{\gamma}_{ij}}$ are now the momenta with respect to Lagrangian \eqref{minioriginal}, instead of the $\pi^{ij}$ of \eqref{miniogeo} used in \eqref{Tmini}. As a result, and  up to a multiplicative constant, there is a difference between of the two functions of a factor $T/T_Y\sim \sqrt{\gamma} $. This simple relation between the two however is derived in the context of pure gravity, no matter fields are considered. In the latter case, both the mini-superspace metric and the potential would change to contain the additional degrees of freedom. This can affect the form, or even the existence, of the homothetic vector.

\subsection{The FLRW example}

To illustrate the application of the previously derived relations to a particular cosmological system, let us briefly consider the case of a FLRW universe with non-zero spatial curvature and with a massless scalar field minimally coupled to gravity. The mini-superspace Lagrangian which successfully reproduces Friedmann's equations is
\begin{equation}
  L = \mathcal{V}_0 \left( -\frac{3 a \dot{a}^2}{\kappa  N}+\frac{a^3 \dot{\phi}^2}{N} + N \frac{3 k a }{\kappa } \right),
\end{equation} 
where $k$ stands for the spatial curvature. The equivalent geodesic Lagrangian is provided under the change $N = \kappa  \mathcal{V}_0 n/(6 a)$ and it reads
\begin{equation} \label{geoLagFLRW}
  L_g = \frac{6}{n} \left(-\frac{3 a^2 \dot{a}^2}{\kappa^2}+\frac{a^4 \dot{\phi}^2}{\kappa} \right)+\frac{n}{2} k \mathcal{V}_0^2.
\end{equation}
The classical solution of the corresponding Euler-Lagrange equations can be easily obtained in the time gauge $n=1$ (note that this is different from the cosmic time gauge $N=1$) and it reads
\begin{align} \label{solFLRW}
  \phi(t) & = \frac{\sqrt{3}}{2 \sqrt{\kappa }} \tanh^{-1}\left(\frac{\kappa  \sqrt{k}\mathcal{V}_0}{\phi_0} t \right), \\
  a(t) & = \frac{1}{\sqrt{3}} \left( \phi_0^2-k\, \kappa ^2 t^2 \mathcal{V}_0^2\right)^{\frac{1}{4}} ,
\end{align}
where $\phi_0$ is a constant of integration.

The mini-superspace metric we read from Lagrangian \eqref{geoLagFLRW} is
\begin{equation}
 G_{\mu\nu} = \left(
\begin{array}{cc}
 -\frac{36 a^2}{\kappa ^2} & 0 \\
 0 & \frac{12 a^4}{\kappa } \\
\end{array}
\right),
\end{equation}
and it has the homothetic vector $\xi = \frac{a}{2} \frac{\partial}{\partial a}$, which satisfies Eq. \eqref{homeq} with conformal factor $\omega=1$.

The corresponding Hamiltonian constraint associated to the Lagrangian \eqref{geoLagFLRW} assumes the form
\begin{equation} \label{HFLRW}
  \mathcal{H} = \frac{\kappa  p_{\phi}^2}{24 a^4}-\frac{\kappa ^2 p_a^2}{72 a^2} - \frac{k \mathcal{V}_0^2}{2} \approx 0 ,
\end{equation}
where we have implemented the usual definitions for the momenta $p_a =\frac{\partial L_g}{\partial\dot{a}}$ and $p_\phi =\frac{\partial L_g}{\partial\dot{\phi}}$. With the help of the homothetic vector $\xi$ we define the phase space function
\begin{equation} \label{TFLRW}
  T = \frac{1}{k \mathcal{V}_0^2} \frac{a}{2} p_a = -\frac{18 a^3 \dot{a}}{\kappa ^2 k \mathcal{V}_0^2 n}.
\end{equation}
It is straightforward to check that on mass shell, i.e. when \eqref{solFLRW} are used together with $n=1$, we obtain $T=t$. That is, $T$ becomes equal to the proper time of the geodesic problem. To compare with the York time, which at the gauge $N=1$ is proportional to $T_Y \sim a'/a$ (we use a prime to denote that now the time variable is different), our choice Eq. \eqref{TFLRW} in that same gauge $N=1$ reads $T\sim a^2 a'$, which is in accordance with what we mentioned in the previous section.

It is quite common to also adopt the scalar field $\phi$ as a time variable, since the corresponding WDW equation $\widehat{\mathcal{H}}\Psi=0$ is given by
\begin{equation} \label{WDWFLRW}
  \frac{\kappa ^2 }{72 a^2}\frac{\partial^2 \Psi}{\partial a^2}  +\frac{\kappa ^2 }{72 a^3} \frac{\partial \Psi}{\partial a} - \frac{\kappa}{24 a^4}\frac{\partial^2 \Psi}{\partial \phi^2}-\frac{1}{2} k \mathcal{V}_0^2 \Psi =0,
\end{equation}
and has a solution of the form $\Psi = e^{i\lambda \phi} \psi(a)$, where $\lambda$ a real constant and $\psi(a)$ satisfies a Bessel equation. The imaginary exponential of $\phi$ in the wave function reminds of the way the time variable enters the wave function as obtained from a time dependent Scr\"odinger equation. In reality, any choice of a monotonic combination of internal degrees of freedom can serve as a valid time variable to model the evolution of the system. It is however of special interest, the fact that the $T$ defined in Eq. \eqref{TFLRW} is canonically conjugate to the Hamiltonian. According to definition \eqref{timeop}, the quantum analogue of \eqref{TFLRW} is
\begin{equation}
  \widehat{T}= -\frac{\ima \hbar }{k \mathcal{V}_0^2} \left( \frac{a}{2} \frac{\partial}{\partial a} + 1 \right).
\end{equation}
It can be easily verified that for any $\Psi$ satisfying \eqref{WDWFLRW}, the equality $[\widehat{T},\widehat{\mathcal{H}}]\Psi =\ima h \Psi$ holds.  

In a crude attempt to derive a Scr\"odinger equation for $T$ variable of \eqref{TFLRW}, we can introduce the canonical transformation $(a,p_a)\rightarrow(T,p_T)$ with
\begin{equation} \label{canonical} 
  \frac{1}{2 k \mathcal{V}_0^2} a\, p_a =T, \quad  -2 k \mathcal{V}_0^2 \ln a= p_T.
\end{equation}
The latter transforms the Hamiltonian constraint to
\begin{equation} \label{Happrox1} 
  \frac{\kappa }{24}  p_{\phi}^2 -\frac{\kappa ^2 k^2 \mathcal{V}_0^4}{18} T^2 - \frac{k \mathcal{V}_0^2}{2}  e^{-\frac{2 p_T}{k \mathcal{V}_0^2}} \approx 0.
\end{equation}
In order to arrive to the above expression we took advantage of the fact that the Hamiltonian constraint is zero, and multiplied with an overall factor $\exp(-2 p_T/(k \mathcal{V}_0^2))$ so as to decouple $p_T$ from the kinetic term. At this point we introduce the following approximation: we consider a slowly varying scale factor around the value $a\sim 1$. Due to the scaling invariance of the Friedmann system we can always normalize the scale factor to unity at any given time, here we do show at a region where $a$ varies slowly, so that from \eqref{canonical} we may demand $p_T<<1$ and $T^2\sim \dot{a}^2\sim 0$ at the same time. Under this approximation \eqref{Happrox1} reduces to
\begin{equation}
   \frac{\kappa }{24}  p_{\phi}^2 - \frac{k \mathcal{V}_0^2}{2}  + p_T \approx 0,
\end{equation}  
which involves a linear term and $p_T$ and can be seen as the classical origin of a Scr\"odinger equation.

\section{Conclusion}

Using the classical correspondence of cosmological and geodesic Lagrangians, we introduce a time variable based on a phase-space function which is related to the proper time. We defined the corresponding operator in the process of canonical quantization and demonstrated that this quantity is canonically conjugate to the Hamiltonian. 

Parametrization invariant systems, due to their zero Hamiltonians, are often accused of not offering time evolution in what regards the quantum states. Here, we uncover the quite striking result that the parametrization invariance itself remedies the lack of a quantum time operator in the usual QM. This suggests a certain time-energy uncertainty relation, although the later may be open to different interpretations, especially in the case of cosmology. In the particle dynamics description it is clear how to assign the fundamental constants in order to have operators with units of time and energy. When considering cosmological Lagrangians, the constants involved in the procedure may change depending on the matter content and/or the gravitational theory under consideration. So, the interpretation of such a relation requires a study case by case. Even in the process of writing the differential operators corresponding to the momenta, one may need to consider an effective Planck constant which involves certain cosmological parameters \cite{Barrow}. Finally, when considering mini-superspace Lagrangians emerging from pure GR, and for a spatially homogeneous spacetime, we saw that the time variable we introduced bears certain similarities to the York time. In a simple example in FLRW cosmology with a massless scalar field, we investigated under which approximations a Scr\"odinger type of equation could be derived from the introduced time variable.

Time parametrization and the resulting vanishing Hamiltonian appear to be the reasons that allow for such a time operator. In the relativistic case, time is an internal variable and can be identified with a specific operator. In the non-relativistic limit, it appears that it ``decouples'' from the rest of system, becoming an external variable and obscuring its quantum properties.  This investigation paves the way for further future studies with respect to this time variable in various gravitational configurations.

\end{document}